\documentclass[12pt]{iopart}

\usepackage{iopams}
\expandafter\let\csname equation*\endcsname\relax
\expandafter\let\csname endequation*\endcsname\relax
\usepackage{amsmath}
\usepackage{textcomp} 
\usepackage{amsopn}
\usepackage{mathrsfs,amssymb} 
\usepackage{bm}       
\usepackage{booktabs} 
\usepackage{dcolumn}  
\usepackage[printonlyused]{acronym} 
\usepackage[normalem]{ulem} 
\usepackage{verbatim} 
\usepackage{capt-of}
\usepackage{relsize}
\usepackage{xcolor}
\usepackage{multirow}
\usepackage{tabularx}
\usepackage{enumitem}
\usepackage{slashed}
\usepackage{mwe}
\usepackage{subfig}
\usepackage{cancel}
\usepackage{bm}
\usepackage{graphicx,epsfig}
\usepackage{amssymb}
\usepackage{multirow}
\usepackage{booktabs}
\usepackage{siunitx}
\usepackage{dcolumn}     

\newcommand{\pr}{\text{pr}}
\newcommand{\myvec}[1]{{\boldsymbol {#1}}}
\newcommand{\bra}{{\langle}}
\newcommand{\ket}{{\rangle}}

\begin{document}

{

\title{Bayesian analysis of nuclear polarizability corrections to the  Lamb shift of muonic H-atoms and He-ions}

\author{S.~S.~Li Muli$^{1}$, B.~Acharya$^{1,2}$, O.~J.~Hernandez$^{1}$, and S.~Bacca$^{1,3}$}
\address{$^{1}$
Institut f\"ur Kernphysik and PRISMA Cluster of Excellence, Johannes Gutenberg-Universit\"at Mainz, 55128 Mainz, Germany}
\address{$^{2}$ Physics Division, Oak Ridge National Laboratory,
Oak Ridge, TN 37831, USA} 
\address{$^{3}$ Helmholtz-Institut Mainz, Johannes Gutenberg-Universit\"at Mainz, D-55099 Mainz, Germany}

\date{\today}

\begin{abstract}

The extraction of nuclear charge radii from spectroscopy experiments in muonic atoms is currently limited by the large uncertainties associated with the theoretical evaluation of the nuclear polarizability effects. To facilitate  calculations, these polarizability corrections are conventionally expressed as an expansion in a dimensionless parameter $\eta$, which has been argued in previous literature to hold an approximate value of 0.33 in light-nuclear systems. In this work, we check this claim by doing a Bayesian analysis of the nuclear-polarizability corrections to the Lamb shift in $\mu{}^2$H and $\mu{}^3$H atoms and in $\mu{}^3$He$^+$ and $\mu{}^4$He$^+$ ions at various orders in the $\eta$-expansion. Our analysis supports the claim that $\eta\ll1$ in these systems and finds truncation uncertainties that are similar to the previous estimate, the only exception being the truncation uncertainties in the $\mu {}^3$He$^+$ ion, which are found to be larger.

\end{abstract}


\maketitle

}


\section{Introduction}\label{Section: Introduction}

Recent advancements in nuclear theory  have enabled us to build quite a robust picture of the nucleus described in terms of interacting protons and neutrons. 
From the theoretical point of view, today's challenges are, on the one hand, to improve the precision via which we can perform our calculations, and, on the other hand, to extend the suite of observables that we can calculate. From the experimental point of view, nuclear radii are fundamental quantities to measure and challenge theory.
While neutron distributions are elusive but can be experimentally accessed, e.g., via  parity violation electron scattering~\cite{CREX,PREX, Thiel_2019,Hagen2016} or coherent elastic neutrino-nucleus scattering~\cite{CEvNS,Cadeddu,Payne19}, the distribution of the protons, traditionally measured with electron scattering, can be investigated with improved precision by studying
 bound systems composed of a muon and a nucleus. With the muon being roughly 200 times heavier than the electron, the energy spectrum of the bound muon is much more sensitive to the details of the nuclear structure compared to a bound electron, allowing  muonic systems to be an excellent precision probe for nuclear physics.

While several heavy muonic ions were investigated in the past~\cite{Schaller85,Shera89,Piller90,Fricke92},  over the last decade the emphasis has been on measurements of light muonic systems~\cite{Pohl_2010,Antognini_2013,Pohl_2016,Krauth_2021}, which lead to an improved precision in the extracted charged radii.
The extraction of the nuclear charge radii is performed through a precise spectroscopic measurement of the muonic Lamb shift  and accurate theoretical calculations. The Lamb shift, $\delta_{\text{LS}}$, is related to the charge radius, $r_c$, by 
\begin{equation}\label{eq: delta_LS}
    \delta_{\text{LS}} = \delta_{\text{QED}} + r_c \ \delta_{\text{OPE}} + \delta_{\text{TPE}}\,.
\end{equation}
Here, $\delta_{\text{QED}}$ includes all the radiative effects that stem from a quantum electrodynamical (QED) description of the process~\cite{Eides2001},  $\delta_{\text{OPE}}$ accounts for the nuclear finite-size corrections and can be pictured as a one-photon-exchange (OPE) diagram between the muon and the nucleus with a charge form factor insertion in the nuclear current, and lastly $\delta_{\text{TPE}}$ is the two-photon-exchange (TPE) nuclear correction~\cite{Ji_2018}\footnote{We note that Eq.~(\ref{eq: delta_LS}) neglects both corrections arising from the exchange of three or more photons \cite{pachucki_2018} and effects of the weak force.}. By measuring the Lamb shift and computing  the terms on the right-hand side of Eq.~(\ref{eq: delta_LS}), one is able to extract the charge radius of the nucleus under consideration.

The study of light muonic systems within modern nuclear theory enables one  to face both of the above mentioned theoretical challenges. On the one hand, given a value of $r_c$, $\delta_{\text{TPE}}$ can be seen as a new observable to be extracted experimentally \cite{Pohl_2016} and used to test nuclear theory. On the other hand, in order to extract  $r_c$  with high precision, the theoretical calculation of $\delta_{\text{TPE}}$ should be precise and accurate.

Given that the aim is to extract charge radii with the highest possible precision, it is crucial to rigorously assess uncertainties in the theoretical contributions. At present, there is a great interest in improving the precision of the calculation of the $\delta_{\text{TPE}}$ term because it is the dominant source of uncertainty in the extraction of $r_c$ from Eq.~(\ref{eq: delta_LS})~\cite{Ji_2018}. This makes it worthwhile to further investigate the uncertainties that accompany the  $\delta_{\text{TPE}}$ calculations.

\begin{figure}[htb]
\begin{center}
\includegraphics[scale=0.12]{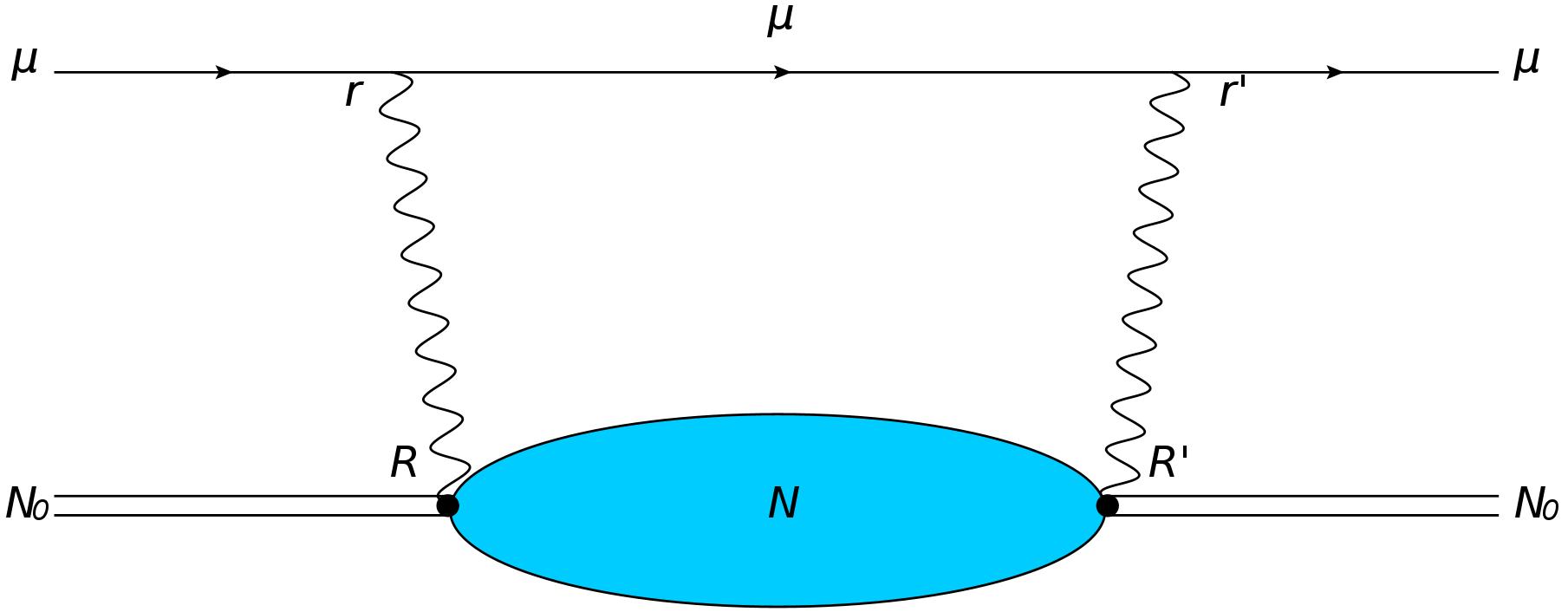}
\end{center}\caption{Diagram of the polarizability correction, where the nucleus is excited between two photon vertexes. In the diagram, $\mu$ represents the muon state, $N_0$ and $N$ are correspondingly the ground and excited nuclear states, while $(r,r')$ and $(R,R')$ are the coordinates of the lepton-photon and nucleus-photon vertex, respectively.}
\label{fig:tpe}
\end{figure}

It is common to split the total two-photon-exchange correction $\delta_{\text{TPE}}$ into an elastic contribution $\delta_{\text{Zem}}$,  which  accounts for corrections where the nucleus remains in the ground state, and a polarizability contribution $\delta_{\text{pol}}$, where the nucleus undergoes transitions from the ground state to its excited states between the photon exchanges, see Fig.~\ref{fig:tpe}. We note that a full calculation of the polarizability correction requires a complete knowledge of all possible excited states of the system.  Although excitation of the single nucleons can  contribute, we neglect them in this work\footnote{These effects have been labelled in Ref.~\cite{Ji_2018} as $\delta_{\text{pol}}^N$ and the total polarizability is a sum of nuclear and pure nucleonic corrections $\delta_{\text{pol}}=\delta_{\text{pol}}^N+\delta_{\text{pol}}^A$.} and restrict our study to the nuclear polarizability correction, which we denote by $\delta^A_{\rm pol}$ to emphasize that we refer to the $A$-body nuclear dynamics \cite{Ji_2018}. In this paper we will analyze only the non-relativistic contributions of the polarizability corrections and, to simplify the notation, still denote them as $\delta^A_{\rm pol}$, as opposed to
$\delta^{A,\rm NR}_{\rm pol}$ which was used in Ref.~\cite{Ji_2018}. Relativistic corrections are small (amounting to 1.5$\%$, 3.7$\%$, 3.6$\%$ and 6.5$\%$~\cite{Ji_2018} in the $\mu{}^2$H, $\mu{}^3$H, $\mu{}^3$He$^+$ and $\mu{}^4$He$^+$ systems, respectively) and are therefore not expected to modify the scaling of the terms in the $\eta$ expansion.

Motivated by the experimental campaign of the CREMA collaboration, in recent times different groups calculated the nuclear polarizability effects for different muonic systems. Calculations for $\mu^2$H were addressed in Refs.~\cite{Pachucki_2011,Friar_2013,Hernandez_2014,Pachucki_2015,Hernandez18}, results for heavier muonic systems systems such as $\mu^3$H, $\mu^3$He$^+$, $\mu^4$He$^+$, $\mu^6$Li$^{2+}$ and $\mu^7$Li$^{2+}$ can be found in Refs.~\cite{Ji_2013,Nevo_Dinur_2016,Muli:2019mkw}, while excellent reviews and intermediate status report are found in Refs.~\cite{Hernandez_2016,Hernandez2017,Ji_2018}. In these works, $\delta^A_{\rm pol}$ values for different systems were computed by performing a Taylor expansion in the operator $\eta = \sqrt{2m_r\omega_N} |\myvec{R}-\myvec{R}'|$, where $m_r$ is the reduced mass of the muon-nuclear system, $\omega_N$ is the nuclear excitation energy and $|\myvec{R}-\myvec{R}'|$ can be pictured as the ``virtual" distance that a nucleon travels inside the nucleus during the TPE process, see Fig.~\ref{fig:tpe}. In Ref.~\cite{Ji_2018}, $\eta$ was estimated to be approximately 0.33 for light nuclear systems by invoking the uncertainty principle to relate the operator $|\myvec{R}-\myvec{R}'|$ to the momentum scale corresponding to the excitation energy $\omega_N$.
In this work, we put for the first time this claim to the testbench by performing a Bayesian analysis of the data of the nuclear polarizability corrections to the Lamb shift in $\mu ^2$H and $\mu ^3$H atoms and in $\mu ^3$He$^+$ and $\mu ^4$He$^+$ ions. Making different reasonable choices for the Bayesian priors, we derive probability distributions for values of $\eta$ and compute the uncertainties stemming from the truncation of the expansion.
For this, we employ tools originally developed to study the truncation errors in nuclear effective field theories~\cite{Furnstahl_2015}. 

The paper is structured as follows. In Section 2, we review the theoretical framework for the $\eta$-expansion as well as the Bayesian formalism. In Section 3, we present the results of our statistical analysis and Section 4 is reserved for the concluding remarks.

\section{Theoretical framework}\label{Section: theo}

In this section we introduce the $\eta$-expansion method and develop the tools used in the statistical analysis of the polarizability data. Following Ref.~\cite{Ji_2018}, we write the non-relativistic polarizability contribution to the Lamb shift in muonic ions as
\begin{equation}
\delta^{A}_{\rm pol} = \sum_{N\neq N_0} \int d^3R \ d^3R' \rho^p_N(\myvec{R})W(\myvec{R},\myvec{R}',\omega_N)\rho^p_N(\myvec{R}')\,, \label{eq: delta_A_pol with W}
\end{equation}
where $\rho^p_N(\myvec{R})=\bra N | \frac{1}{Z} \sum_{i=1}^A \delta(\myvec{R}-\myvec{R_i}) \ \hat{e}^{p}_i \ |N_0\ket$ is the proton transition density, $\hat{e}^p_i$ is a proton-projection operator on the (isospin) Hilbert space of the $i$-th nucleon, $N$ and $N_0$ are nuclear quantum numbers with $N_0$ representing the ground state, $\omega_N$ is the excitation energy of the $N$-th nuclear state, $Z$ is the charge number of the nucleus, and $W(\myvec{R},\myvec{R}',\omega_N)$ is the lepton matrix element. After carrying out the integral over the exchanged-momentum $q$,  $W(\myvec{R},\myvec{R}',\omega_N)$ can be written in terms of $\eta$ as
\begin{equation}
W(\myvec{R},\myvec{R}',\omega_N) = -\frac{\pi}{m^2_r} \ (Z \alpha)^2 \ \phi^2(0) \ \Bigl(\frac{2m_r}{\omega_N}\Bigr)^{\frac{3}{2}}\frac{1}{\eta}\left( e^{-\eta} -1 + \eta - \frac{1}{2}\eta^2 \right).
\label{eq: lept_ME}
\end{equation}
Here, 
$\alpha$ is the fine-structure constant and $\phi^2(0)$ is the value of the muonic 2S-state wave function at the origin.
Next, we expand Eq.~(\ref{eq: lept_ME}) in a Taylor series over $\eta$ obtaining

\begin{equation}
    W(\myvec{R},\myvec{R}',\omega_N) = \frac{\pi}{6 m_r} \ (Z \alpha)^2 \ \phi^2(0) \ \Bigl(\frac{2m_r}{\omega_N}\Bigr)^{\frac{3}{2}}\Bigl[ \ \eta^2 - \frac{1}{4} \ \eta^3 + \frac{1}{20} \ \eta^4 + \ ... \ \Bigr]\,.
    \label{eq: lept_ME_2}
\end{equation}
Inserting Eq.~(\ref{eq: lept_ME_2}) into Eq.~(\ref{eq: delta_A_pol with W}), we obtain the polarizability correction as an expansion in $\eta$ 
\begin{eqnarray}
\label{eq: delta_pol_expansion}
    \delta^{A}_{\rm pol} &= \frac{\pi}{6 m_r} \ (Z \alpha)^2 \ \phi^2(0) \ (2m_r)^{\frac{3}{2}} \sum_{N\neq N_0} \Bigl(\frac{1}{\omega_N}\Bigr)^{\frac{3}{2}} \times\\
    \nonumber
    & \times  \int d^3R \ d^3R' \rho^p_N(\myvec{R}) \Bigl[ \ \eta^2 - \frac{1}{4} \ \eta^3 + \frac{1}{20} \ \eta^4 + \ ... \ \Bigr] \rho^p_N(\myvec{R}') \\
    \nonumber
&= D_2 + D_3 + D_4 + ...     \, ,
\end{eqnarray}
where the index in the ``$D$" terms in the bottom line  indicates the power of $\eta$ entering each term in the above line.
Note that in Ref.~\cite{Ji_2018} the leading ($D_2$), subleading ($D_3$) and sub-sub-leading ($D_4$) contributions were labelled as $\delta^{(0)}$, $\delta^{(1)}$ and $\delta^{(2)}$, respectively.

We now assume that $\delta^{A}_{\rm pol}$ can be expressed as a power law in $\eta$ 
\begin{equation}
\delta^{A}_{\rm pol} 
= X_{\text{ref}} \ \Bigl[ \ c_2 \ \eta^2 + c_3 \ \eta^3 + c_4 \ \eta^4 + ... \ \Bigr]\,,
\label{eq: delta_pol_model}
\end{equation}
where $c_i$ are dimensionless coefficients that  may depend on mass number $A$ and charge number $Z$, and $X_{\rm ref}$ is the natural scale of $\delta^{A}_{\rm pol}$. Starting with this assumption we perform a Bayesian analysis of the convergence of the $\eta$-expansion. We use the data in Table~\ref{tab: data} taken from Ref.~\cite{Ji_2018} which correspond to calculations in chiral effective field theory ($\chi$-EFT) with an Hamiltonian that includes a two-body force at next-to-next-to-next-to-leading order~\cite{Entem_2003} and a three-body force at next-to-next-to-leading order~\cite{Na07}. To ensure that the coefficients $ c_i$ are dimensionless and natural, i.e. $\mathcal{O}(1)$, the reference $X_{\text{ref}}$ should have the same dimensions and order of magnitude as the analyzed observable. We take as $X_{\text{ref}}$ the $\delta_{\rm TPE}$ value computed in Ref.~\cite{Ji_2018} with the phenomenological AV18+UIX interaction~\cite{AV18,PuP95}. 

We notice that going from Eq.~(\ref{eq: delta_pol_expansion}) to Eq.~(\ref{eq: delta_pol_model}) is analogous to what has been extensively done in  $\chi$-EFT. There, the nuclear Hamiltonian is expanded in powers of a dimensionless ratio $Q/\Lambda$ where $Q$ is the  typical low momentum scale of the process under study and $\Lambda$ is the breakdown scale. The calculated low-energy observables are then assumed to inherit this $Q/\Lambda$ expansion and uncertainties are quantified by studying their order-by-order convergence~\cite{Furnstahl_2015,Hernandez18,Ekstrom2020,Epelbaum2020,Acharya:2021lrv}.

As in $\chi$-EFT, the framework of Eq.~(\ref{eq: delta_pol_expansion}) and (\ref{eq: delta_pol_model}) is only useful if the expansion parameter $\eta$ has a value smaller than 1. 
In particular, since the higher-order terms become progressively more complicated to calculate~\cite{Ji_2018}, the practical utility of this expansion depends on a fast convergence  so that meaningful estimates can be obtained by truncating the series at low orders.


\begin{table}[h!]
\begin{center}
\begin{tabular}{ c | c| c| c | c |c }
                &In notation Ref.~\cite{Ji_2018} & $^2$H & $^3$H & $^3$He & $^4$He \\
  \hline			
  $D_2$ & $\delta^{(0)}_{D1}$ &-1.912 & -0.7848 & -6.633 & -4.701 \\
  $D_3$ & $\delta^{(1)}_{R3}+\delta^{(1)}_{Z3}$ &0.359 & 0.1844 & -0.384 & 0.809\\
  $D_4$ & $\delta^{(2)}_{Q}+\delta^{(2)}_{D1D3} + \delta^{(2)}_{R^2} $ &-0.037 & -0.0247 & 0.83 & 0.101\\
  $\sum_{k=2}^4 D_k$ & -& -1.590 & -0.6251 & -6.187 & -3.791 \\
  $\text{X}_{\text{ref}}$ &- & -1.664 & -0.6986 & -14.564 & -8.220 \\
  \hline  
\end{tabular}
\caption{The nuclear polarizability correction to the Lamb shift (in  meV) for several  light muonic atoms at various orders in the $\eta$-expansion. The reference $\text{X}_{\text{ref}}$ are the $\delta_{\rm TPE}$ values evaluated using the AV18+UIX interaction. Data taken from Ref.~\cite{Ji_2018}.}
\label{tab: data}
\end{center}
\end{table}


Next, we introduce the Bayesian method that allows us to compute the probability distribution for the value of $\eta$ and to obtain the truncation error in the $\eta$-expansion therefrom. Given a data set $\{ D_m,...,D_k \}$, Bayes' theorem expresses the posterior probability distribution of $\eta$ as 
\begin{equation}
\pr(\eta| D_m,...,D_k) =  \frac{\pr(D_m,...,D_k|\eta) \ \pr(\eta)}{\pr(D_m,...,D_k)}\,.
\label{eq: bayes}
\end{equation}
Here, $\pr(D_m,\ldots,D_k|\eta)$ is known as the likelihood function and represents the probability of obtaining the data set given a particular value of $\eta$. The prior (``a priori") probability distribution $\pr(\eta)$ should summarize the information we have about $\eta$ before analyzing the data set. The denominator $\pr(D_m,...,D_k)$ is the marginal likelihood, which, being independent of $\eta$, can be taken as a normalization constant for our purpose. We make the assumption that the dimensionless coefficients $c_i$, which are related to the data set $\lbrace D_m,\ldots, D_k\rbrace$ by $D_i = c_i \, \eta^i$, are uncorrelated among each other and their scale is set by a single scale parameter $\bar{c}$ through a probability distribution $\pr(c_i|\bar{c})$ for $i=m,...,k$. We then eliminate the dependence on this ``nuisance" parameter $\bar{c}$ by marginalizing over it as
\begin{eqnarray}
 \nonumber
\pr(D_m,...,D_k|\eta) = \int dc_m...dc_k \ \pr(D_m,...,D_k|c_m,...,c_k,\eta)\pr(c_m,...,c_k|\eta) \\
 \label{eq: likelihood}
= \int d\bar{c}\ dc_m...dc_k \ \pr(D_m,...,D_k|c_m,...,c_k,\eta) \ \pr(c_m,...,c_k|\bar{c},\eta) \ \pr(\bar{c}|\eta) \,.
\end{eqnarray}
It follows from Eq.~(\ref{eq: delta_pol_model}) and from the assumption that the dimensionless coefficients $c_i$ are uncorrelated that 
\begin{eqnarray}
\nonumber
 \pr(D_m,...,D_k|c_m,...,c_k,\eta) = \prod_{i=m}^k \pr(D_i | c_i, \eta) = \prod_{i=m}^k \delta(D_i - \eta^i c_i) \\ 
	 = \prod_{i=m}^k \frac{1}{\eta^i} \delta(D_i/\eta^i-c_i)= \frac{1}{\eta^{(k-m+1)(k+m)/2}} \prod_{i=m}^k \delta(D_i/\eta^i-c_i)\,.
	\end{eqnarray}
Using the delta functions to  analytically perform the integrals over the dimensionless coefficients in Eq.~(\ref{eq: likelihood}), we obtain
\begin{equation}
\pr(D_m,...,D_k|\eta) = \frac{1}{\eta^{(k-m+1)(k+m)/2}}  \int \pr\Bigl(c_m=\frac{D_m}{\eta^m} \ ,..., \ c_k=\frac{D_k}{\eta^k} \ \Bigl| \ \bar{c},\eta\Bigr) \ \pr(\bar{c}) \ d\bar{c} \ ,
\end{equation}
where we further assumed that the scale-parameter $\bar{c}$ is independent of $\eta$. The posterior (``a posteriori") $\pr(\eta|D_m,...,D_k)$ can then be written as
\begin{equation}
\pr(\eta|D_m,...,D_k) = \frac{\mathcal{N}}{\eta^{(k-m+1)(k+m)/2}} \int \pr\Bigl(c_m=\frac{D_m}{\eta^m} \ ,..., \ c_k=\frac{D_k}{\eta^k} \ \Bigl| \ \bar{c},\eta\Bigr) \ \pr(\bar{c}) \ \pr(\eta) \ d\bar{c} \ ,
\label{eq: eta_Dm_Dk}
\end{equation}
where the normalization constant is $\mathcal{N}=\pr(D_m,...,D_k)^{-1}$. 
\\

It was argued in Ref.~\cite{Ji_2018} that the parameter $\eta$ has an approximate value of 0.33 for light-nuclear systems, and has, at most, a weak dependence on the exact system at hand. It is then interesting to explore the consequences of {\it combining the data sets} $\lbrace D_m,\ldots, D_k\rbrace$ for different systems under the assumptions that they correspond to the same underlying probability distribution for $\eta$ and for the expansion coefficients $c_i$. To this end, we work with a  new data set $\{ D_m^0  , ...  , D_k^n \} \equiv \{ D_m^0 \ , ... , D_k^0 \ , ... , D_m^n \ , ...  , D_k^n \}$ where $\lbrace D_m^0,\ldots, D_k^0\rbrace$ is the data set of $\mu^2$H, $\lbrace D_m^1,\ldots, D_k^1\rbrace$ is that of $\mu^3$H, $\lbrace D_m^2,\ldots, D_k^2\rbrace$ that of  $\mu^3$He$^+$ and finally $\lbrace D_m^3,\ldots, D_k^3\rbrace$ is that of  $\mu^4$He$^+$. Proceeding as before, the posterior for $\eta$ can be written as
\begin{eqnarray}
\label{eq: eta_Dm_Dk_combined}
\pr(\eta|D^0_m,...,D^n_k) = \\
\nonumber
\frac{\mathcal{N}}{\eta^{(n+1)(k-m+1)(k+m)/2}} \int \pr\Bigl(c^0_m=\frac{D^0_m}{\eta^m} \ ,..., \ c^n_k=\frac{D^n_k}{\eta^k} \ \Bigl| \ \bar{c},\eta\Bigr) \ \pr(\bar{c}) \ \pr(\eta) \ d\bar{c} \ ,
\end{eqnarray}
where the normalization constant $\mathcal{N}$ now denotes  $\pr(D^0_m,...,D^n_k)^{-1}$.

Next, we want to find a probability distribution for the truncation error of the $\eta$-expansion  of Eq.~(\ref{eq: delta_pol_model}), $\Delta_k^{(1)}$,  in the approximation that this is dominated by the first omitted term,  $\Delta_k^{(1)}\sim c_{k+1} \ \eta^{k+1} $. Below, we give a brief overview of the Bayesian methodology for quantification of truncation errors and refer the reader to  Ref.~\cite{Furnstahl_2015} for further details.

Given the data set $\{ D_m,...,D_k\}$, we begin by expressing the Bayesian posterior for the truncation error at a given value of $\eta$, $\pr(\Delta^{(1)}_k \ | \ D_m, ..., D_k,\eta)$, as an integrated likelihood  over the first omitted coefficient $c_{k+1}$, 
\begin{eqnarray}
\nonumber
\pr(\Delta^{(1)}_k \ | \ D_m, ..., D_k,\eta) = \\
\nonumber
=\int d c_{k+1} \ \pr(\Delta^{(1)}_k \ | \ D_m,...,D_k,\eta,c_{k+1}) \ \pr(c_{k+1} \ | \ D_m,...,D_k, \eta) \\
\label{eq: trunc_bef_cbar}
= \int d c_{k+1} \ \delta(\Delta_k^{(1)}-c_{k+1} \ \eta^{k+1}) \ \pr(c_{k+1} \ | \ D_m,...,D_k, \eta) \\
\nonumber
= \frac{1}{\eta^{k+1}} \ \pr \bigl(c_{k+1}=\frac{\Delta^{(1)}_k}{\eta^{k+1}} \ \bigr| \ D_m,...,D_k, \eta \bigr)\,.
\end{eqnarray}
We introduce again the scale parameter $\bar{c}$ by marginalizing over it
\begin{equation}
\pr(\Delta^{(1)}_k \ | \ D_m, ..., D_k,\eta) = \frac{1}{\eta^{k+1}} \int d\bar{c} \ \pr \bigl(c_{k+1}=\frac{\Delta^{(1)}_k}{\eta^{k+1}} \ \bigr| \ D_m,...,D_k, \eta, \bar{c} \bigr) \ \pr(\bar{c} \ | \ D_m,...,D_k, \eta) \,.
\label{eq: trunc_marg_cbar}
\end{equation}
Applying Bayes' theorem on the last term of Eq.~(\ref{eq: trunc_marg_cbar}), we arrive at the expression
\begin{equation}
\pr(\Delta^{(1)}_k| \ D_m, ... , D_k ,\eta) = \frac{\int d\bar{c} \ \pr \bigl( c_{k+1}=\frac{\Delta_k^{(1)}}{\eta^{k+1}} \ \bigr| \ \bar{c} \ \bigr) \ \Bigl[ \ \prod\limits_{i=m}^{k} \pr\bigl( c_i = \frac{D_i}{\eta^i} \ \bigr| \ \bar{c} \ \bigr) \Bigr] \ \pr(\bar{c})}{ \eta^{k+1} \int d\bar{c} \ \Bigl[ \ \prod\limits_{i=m}^{k} \pr\bigl( c_i = \frac{D_i}{\eta^i} \ \bigr| \ \bar{c} \ \bigr) \Bigr] \pr(\bar{c})}.
\label{eq: trunc_prob_final}
\end{equation}
We now marginalize over the expansion parameter $\eta$ to obtain 
\begin{equation}
\pr(\Delta^{(1)}_k| \ D_m, ... , D_k ) = \int d\eta \ \pr(\Delta^{(1)}_k| \ D_m, ... , D_k, \eta ) \ \pr(\eta| \ D_m,...,D_k)\,,
\label{eq: trunc_prob_final_marg}
\end{equation}
with $\pr(\eta| \ D_m,...,D_k)$ given by Eq.~(\ref{eq: eta_Dm_Dk}).

The generalization for the {\it combined data set} $\{ D_m^0 \ , ... , D_k^n \}$ is straightforward and yields
\begin{equation}
\pr(\Delta^{(1)}_k| \ D^0_m, ... , D^n_k ,\eta) = \frac{\int d\bar{c} \ \pr \bigl( c_{k+1}=\frac{\Delta_k^{(1)}}{\eta^{k+1}} \ \bigr| \ \bar{c} \ \bigr) \ \Bigl[ \ \prod\limits_{i=m}^{k} \prod\limits_{j=0}^{n} \pr\bigl( c^j_i = \frac{D^j_i}{\eta^i} \ \bigr| \ \bar{c} \ \bigr) \Bigr] \ \pr(\bar{c})}{ \eta^{k+1} \int d\bar{c} \ \Bigl[ \ \prod\limits_{i=m}^{k} \prod\limits_{j=0}^{n} \pr\bigl( c^j_i = \frac{D^j_i}{\eta^i} \ \bigr| \ \bar{c} \ \bigr) \Bigr] \pr(\bar{c})}\,.
\label{eq: trunc_prob_final_combined}
\end{equation}
After marginalization over $\eta$ we obtain the posterior distribution of the truncation error, 
\begin{align}
\pr(\Delta^{(1)}_k| \ D^0_m, ... , D^n_k ) = \int d\eta \ \pr(\Delta^{(1)}_k| \ D^0_m, ... , D^n_k, \eta ) \ \pr(\eta| \ D^0_m,...,D^n_k)\,,
\end{align}
with $\pr(\eta| \ D^0_m,...,D^n_k)$ given by Eq.~(\ref{eq: eta_Dm_Dk_combined}).

\subsection{Priors}

To proceed with the statistical analysis, we need to specify the prior probability distributions, $\pr(\bar{c})$, $\pr(c_i | \bar{c})$ and $\pr(\eta)$. 
Whenever possible, we will be guided by the principle of maximum entropy \cite{Jaynes57} to find the least informative priors given some basic properties of the parameters $c_i$, $\bar{c}$ and $\eta$.
Below we report a few considerations on the choice of the priors:

\begin{itemize}
    \item The parameter $\bar{c}$ sets the overall scale of the dimensionless coefficients $c_i$. Our unbiased expectations about $\bar{c}$ under the only constraint that it is a positive definite quantity can be encoded by adopting the Jeffreys' prior,  $\pr(\bar{c})\propto 1/\bar{c}$~\cite{Jeffrey39}. In order to work with a normalizable distribution, we introduce a slight modification and restrict the range of $\bar{c}$ from a minimum value of $c_<=10^{-4}$ to a maximum value of $c_>=10^{4}$ by multiplying with step functions $\theta(x)$.
    
    \item For $\pr(c_i | \bar{c})$, we test two different reasonable choices. The extent to which the obtained posterior distributions are sensitive to these choices will tell us whether the data set is sufficiently informative to dominate the analysis. Our first choice, labelled prior A, assumes that the magnitude of the dimensionless coefficients can not be larger than $\bar{c}$. The maximum-entropy principle then leads to a uniform distribution in the range $-\bar{c} < c_i <\bar{c}$ with $i=m,...,k$. Our second choice, labelled prior B, is a Gaussian distribution with zero mean and standard deviation $\bar{c}$. It was first adopted in nuclear effective field theories in  Ref.~\cite{Schindler09} and is motivated by the maximum-entropy principle under the assumption of testable information over the means and standard deviations of the dimensionless coefficients.
    \item For $\pr(\eta)$, we first note that this parameter is constrained to hold a positive value.
    In what follows, we assume that the simple estimate of $\eta \approx 0.33$ can be regarded as an information on the mean of $\pr(\eta)$. The exponential distribution is the least-informative prior for a parameter which is positive definite and whose mean is known \cite{Sivia06}. We label this choice as $\alpha_\eta$. To perform a check on how much the conclusions of the statistical analysis are stable with respect to reasonable modifications of the $\eta$-prior, we compare this choice with a $\beta$-distribution, $\beta(a,b)$, which constrains $\eta \in [0,1]$, and label this second prior choice as $\beta_\eta$. The parameters of $\beta_\eta$ are chosen such that the mean falls on $0.33$. There is an infinite set of $\beta$-distributions that satisfy this constraint. We selected $a=3.0$ and $b=6.0$, because it holds a reasonable standard deviation for both the prior and the data to be informative.

\end{itemize}

We list in Table~\ref{tab: priors1} the prior-choices for $\pr(\bar{c})$ and $\pr(c_i|\bar{c})$, while the prior choices for $\pr(\eta)$ are listed in Table~\ref{tab: priors2}.
\begin{table}[h!]
\begin{center}
\begin{tabular}{ c | c| c }
 Priors & $\pr(c_i|\bar{c})$     & $\pr(\bar{c})$  \\ \hline
 A & $\frac{1}{2\bar{c}}\theta(\bar{c}-|c_i|)$ & $\frac{1}{\text{ln}(\bar{c}_>/\bar{c}_<) \bar{c} }\theta(\bar{c}-\bar{c}_<) \theta(\bar{c}_>-\bar{c})$ \\ 
 B & $\frac{1}{\sqrt{2\pi}\bar{c}}\text{exp}\left(-\frac{c^2_i}{2\bar{c}^2}\right)$ &  $\frac{1}{\text{ln}(\bar{c}_>/\bar{c}_<) \bar{c} }\theta(\bar{c}-\bar{c}_<) \theta(\bar{c}_>-\bar{c})$
\end{tabular}
\caption{Prior choices for the probability density distributions of the scale parameter $\bar{c}$ and for the dimensionless coefficients $c_i$.}
\label{tab: priors1}
\end{center}
\end{table}

\begin{table}[h]
\begin{center}
\begin{tabular}{ c | c}
 Priors & $\pr(\eta)$     \\ \hline
$\alpha_\eta$ & $\frac{1}{\lambda}$ $\exp\bigl(\frac{\eta}{\lambda}\bigr)$ \\ 
$\beta_\eta$ & $\beta(a,b)$
\end{tabular}
\caption{Prior choices for the probability density distribution of the $\eta$ expansion parameter. Following the prediction in Ref.~\cite{Ji_2013,Ji_2018} the mean of the exponential distribution is $\lambda=0.33$, while for the $\beta$-distribution we take $a=3.0$ and $b=6.0$. }
\label{tab: priors2}
\end{center}
\end{table}
 
\newpage
\section{Results}\label{Section: results}
Our first aim is to obtain posterior distributions for values of $\eta$ and check whether they depart from the priors motivated by the estimate of $\eta \sim 0.33$. 

\begin{figure}[htb]
\begin{center}
\includegraphics[scale=0.9]{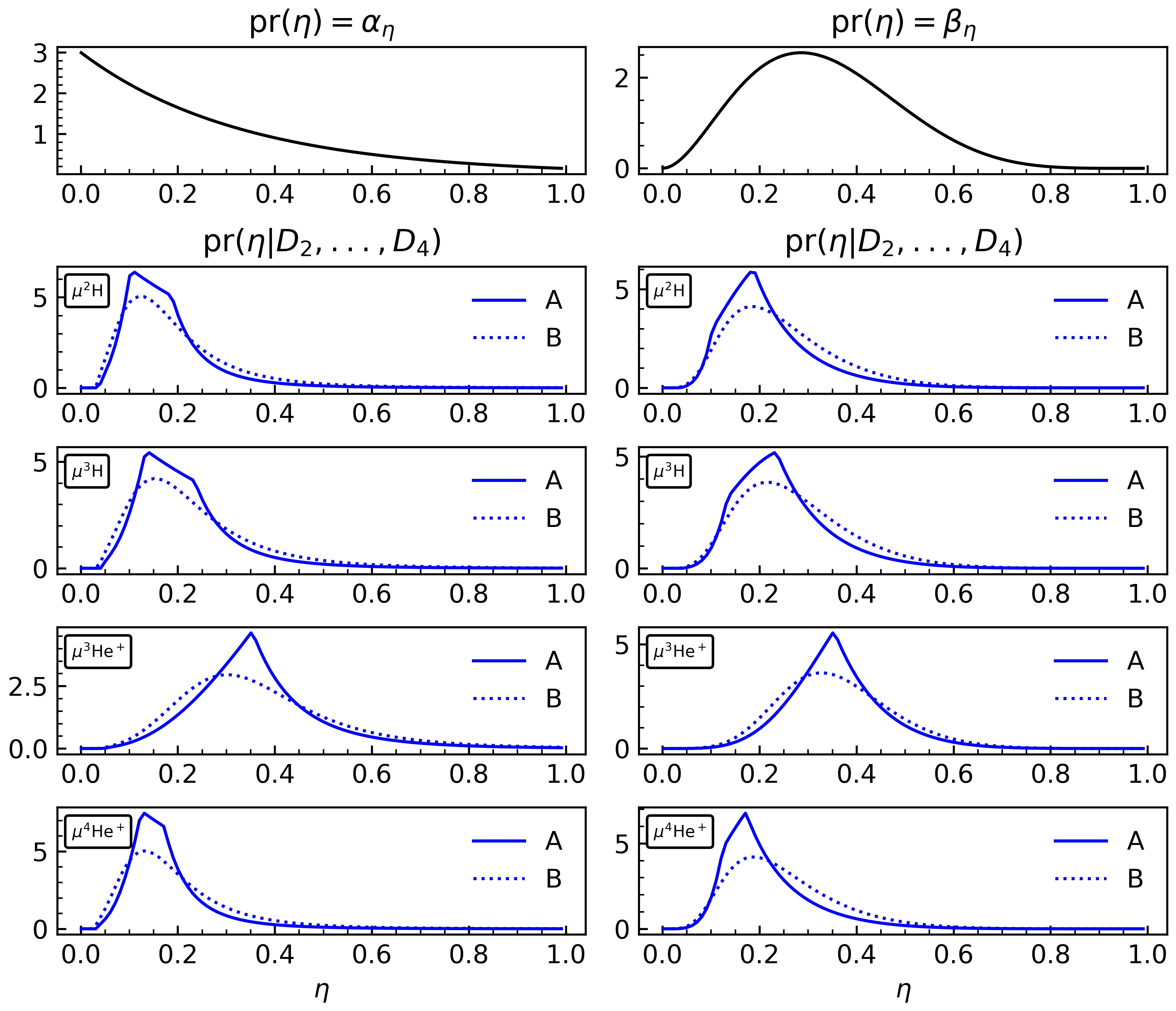}
\end{center}
\caption{Analysis for the {\it separate} muonic systems.(Upper panel)  Prior probability distributions $\pr(\eta)$:  $\alpha_\eta$ choice (left) and $\beta_\eta$ choice (right).
(Lower panels) Corresponding posterior probability distributions  $\pr(\eta| \ D^0_m,...,D^n_k)$ 
for the choice of $\pr(\eta)=\alpha_\eta$ (left) and $\pr(\eta)=\beta_\eta$ (right) and
two given choices (A and B) of priors for $\pr(\bar{c})$ and  $\pr(c_i|\bar{c})$.  
}
\label{fig: Eta_posteriors_sep}
\end{figure}

First, in Figure~\ref{fig: Eta_posteriors_sep} we show the resulting posterior distributions
for $\pr(\eta| \ D_2,...,D_4)$ 
for different choices of the priors $\pr(\eta)$, $\pr(c_i|\bar{c})$ and $\pr(\bar{c})$, when we do a {\it separate} analysis of each individual muonic system. Overall, the posterior distributions are quite stable under modifications of the priors, suggesting that there is enough information in the data sets to dominate the analysis.
In particular for $\pr(\eta)$, we note that the exponential distribution only assumes $\eta > 0$, whereas the $\beta$-distribution constrains $\eta \in [0,1]$. We find only weak deviations in the results with the two priors. This increases the significance of the assumption that $\eta \in [0,1]$, which in turns means that the expansion in Eq.~(\ref{eq: delta_pol_model}) converges. Our Bayesian analysis indicates that the most likely $\eta$-value is smaller than the estimate obtained from the uncertainty principle for the $\mu^2$H, $\mu^3$H and $\mu^4$He$^+$ systems, because the maximum-likelihood value of $\eta$ is approximately $0.15$. 

Given that we find considerably smaller maximum-likelihood values of $\eta$ than the 0.33  suggested by the uncertainty principle, one might ask whether there is enough evidence to reject the 0.33 estimate. An answer can be obtained by calculating the probability $\pr(\eta > 0.33 | \ D_2,...,D_4)$ for the corresponding posterior probability distribution shown in Figure~\ref{fig: Eta_posteriors_sep}, which oscillate between a minimum value of $6\%$ (obtained in the $\mu^4$He$^+$ system with prior choices $\text{A},\alpha_\eta$) and a maximum value of $27\%$ (obtained in the $\mu^3$H system with prior choices $\text{B},\beta_\eta$). We conclude that although it is likely that the values of $\eta$ are actually smaller than previously suggested, there is still enough statistical evidence for the $\eta$ estimate based on the  uncertainty principle to be correct. Lastly, for the case of $\mu^3$He$^+$, we get a much closer agreement with the estimated $0.33$, given that the  maximum-likelihood value of $\eta$ is about $0.35$.

\begin{figure}[htb]
\begin{center}
\includegraphics[scale=0.9]{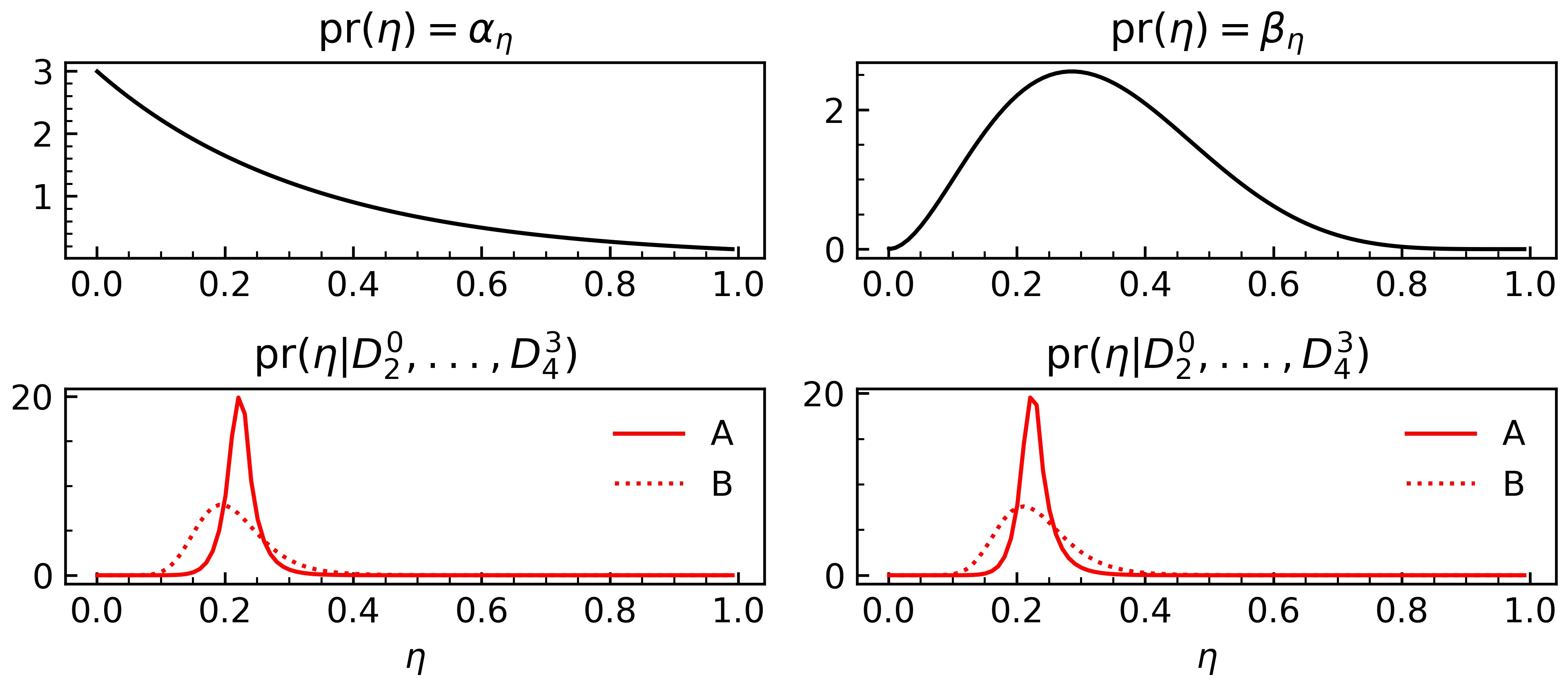}
\end{center}
\caption{Analysis for the {\it combined} muonic systems. 
(Upper panel)  Prior probability distributions $\pr(\eta)$:  $\alpha_\eta$ choice (left) and $\beta_\eta$ choice (right).
(Lower panel) Corresponding posterior probability distributions  $\pr(\eta| \ D_m,...,D_k)$ 
for the choice of $\pr(\eta)=\alpha_\eta$ (left) and $\pr(\eta)=\beta_\eta$ (right) and
two given choices (A and B) of priors for $\pr(\bar{c})$ and  $\pr(c_i|\bar{c})$.
}
\label{fig: Eta_posteriors_comb}
\end{figure}

We then perform a {\it combined} analysis using all the  muonic atoms and show the results in Figure
\ref{fig: Eta_posteriors_comb}. Also in this case we present results for two choices of the priors  $\pr(\eta)$, namely $\alpha_n$ and $\beta_n$, and for the choice A and B of the other two priors, as reported in Table~\ref{tab: priors1}. In this case, we note that the posterior probability distributions  do not depend much on the $\eta$-priors  $\alpha_n$ or $\beta_n$, while choices A and B give quite different results. This last feature occurs because in the process of combining the data sets it is assumed that all the expansion coefficients are drawn from the same underling distribution. Although the assumption seems reasonable for the data sets of $\mu^2$H, $\mu^3$H and $\mu^4$He$^+$, it is  unclear whether the same applies to the $\mu^3$He$^+$ system where the expansion coefficients are -- at fixed $\eta$-values -- systematically larger compared to the other muonic atoms. The effect of the $\mu^3$He$^+$ system in the combined analysis impacts more the results with prior A  than prior B, because calculations based on prior B are less sensitive to extreme values of the coefficients ${c_i}$~\cite{Furnstahl_2015}.
In particular, we find that choice B predicts a lower maximum-likelihood for $\eta$ than choice A, so that Eq.~(\ref{eq: delta_pol_model}) will converge faster. Interestingly, the maximum-likelihoods of $\eta$ in the {\it combined} analysis are  $\eta\sim 0.22$ (A) and $\eta\sim 0.20$ (B), which resemble a sort of ``average" between the individual distributions of the {\it separate} analysis.

\begin{figure}[htb]
\begin{center}
\includegraphics[scale=0.9]{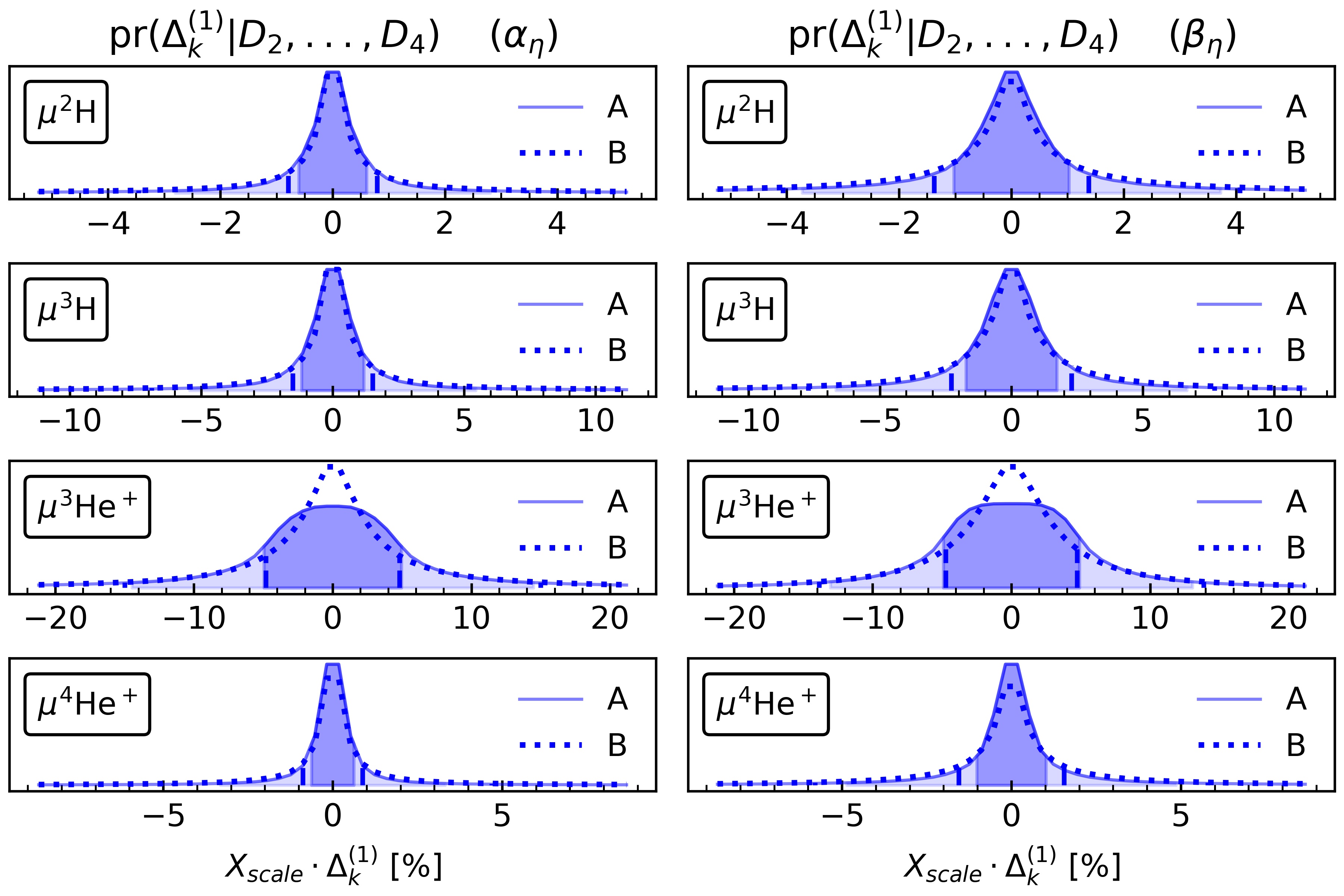}
\end{center}
\caption{Posterior distributions for the 
truncation uncertainties of the $\eta$-expansion in a Bayesian analysis of the {\it separate} muonic systems. Results with $\pr(\eta)=\alpha_{\eta}$ (left) and $\pr(\eta)=\beta_{\eta}$ (right). Solid lines (dotted lines) represent  posterior distributions obtained with choices A (B) for priors of Table~\ref{tab: priors1}. The 68$\%$ and 95$\%$ CI are reported as dark and light shaded areas for the prior choice A, respectively, while for the prior choice B they are reported as vertical dashed lines. The y-axes are given in arbitrary units.}
\label{fig: Delta_posteriors_sep}
\end{figure}

As a next step, we address the evaluation of the truncation errors in the $\eta$-expansion of $\delta_{\text{pol}}^{A}$ with a Bayesian analysis.
Using the posterior distributions $\pr(\eta | D_2,...,D_4)$ of Figure~\ref{fig: Eta_posteriors_sep}, we  calculate the distributions for the truncation uncertainty $\pr(\Delta^{(1)}_k| D_2, ... , D_4 )$ by marginalization as shown in Eq.~(\ref{eq: trunc_prob_final_marg}). We show first the results of the {\it separate} analysis in Figure~\ref{fig: Delta_posteriors_sep} and we tabulate the confidence interval (CI) at the $68 \%$ and $95 \%$ level in Table~\ref{tab: 68_95}.
 Again, the distributions show good stability when the prior choices get modified and we find a fairly good agreement between the $68 \%$ CI uncertainties of this work and the estimates in Refs.~\cite{Ji_2018} as shown in  Table \ref{tab: 68_95}, with the exception of the truncation uncertainties in the $\mu^3$He$^+$ system, which we found to be roughly a factor of 4 larger. In the specific case of this muonic atom, we see that choices A and B yield different posterior distributions but similar $68 \%$ CI. We note that the $95 \%$ CIs in Table~\ref{tab: 68_95} are much larger than twice the size of the corresponding $68 \%$ CIs for all systems and all prior choices, reflecting the large tails of the posterior distributions compared to the Gaussian (see Figure~\ref{fig: Delta_posteriors_sep}).

\begin{figure}[htb]
\begin{center}
\includegraphics[scale=0.9]{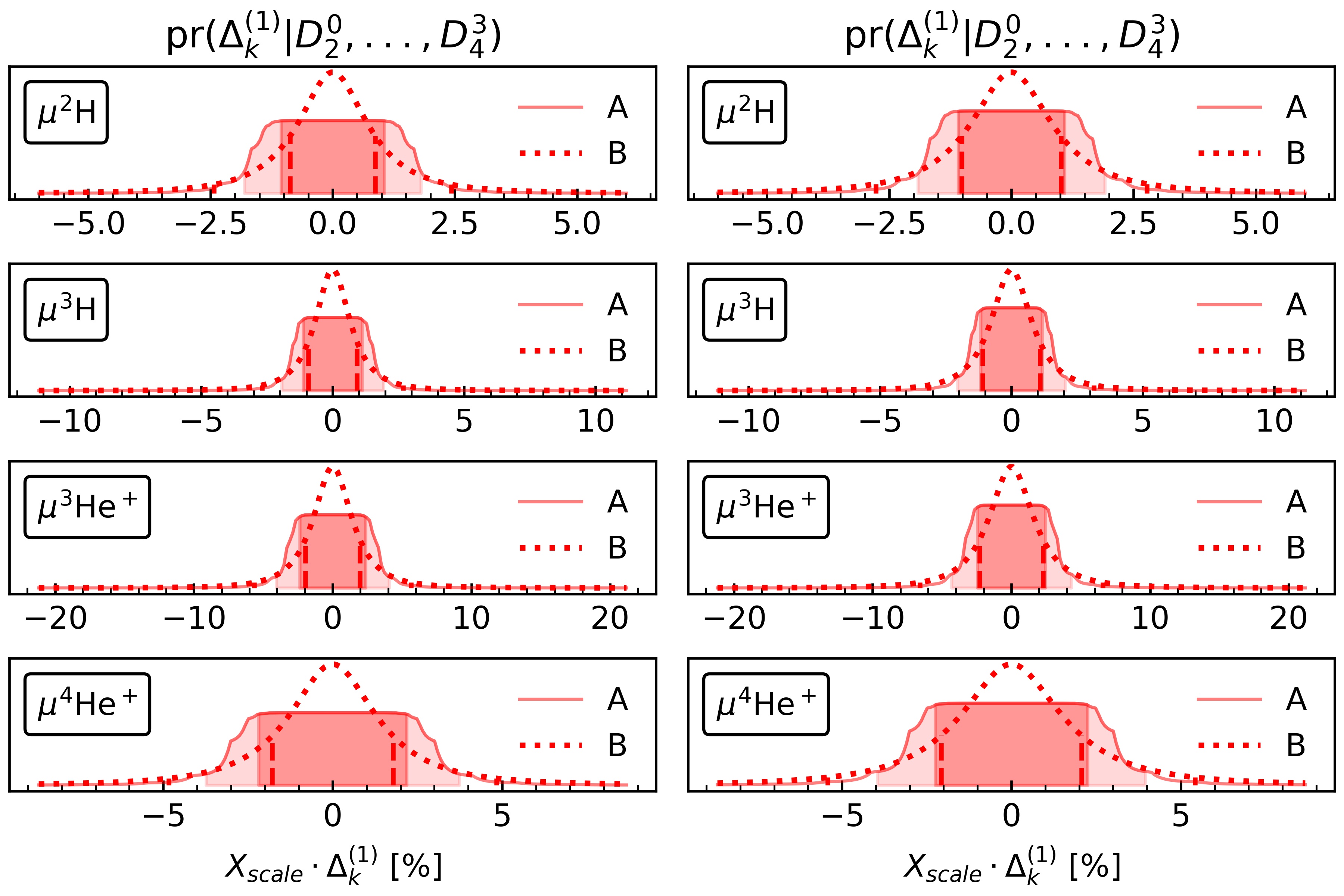}
\end{center}
\caption{ Posterior distributions for the leading order truncation uncertainties of the $\eta$-expansion in a Bayesian analysis for the {\it combined} muonic systems. Results with $\pr(\eta)=\alpha_{\eta}$ (left) and $\pr(\eta)=\beta_{\eta}$ (right). Solid lines represent  posterior distributions obtained with choices A for priors $c_k$'s and $\bar{c}$, while  dotted  lines correspond to the choice B. The 68$\%$ and 95$\%$ CI are reported as dark and light shaded areas for the prior choice A, respectively, while for the prior choice B they are reported as vertical dashed lines. The y-axes are given in arbitrary units.}
\label{fig: Delta_posteriors_allcomb}
\end{figure}

When performing the analogous  analysis of the {\it combined} data, reported in Figure~\ref{fig: Delta_posteriors_allcomb} and in Table~\ref{tab: 68_95_comb}, we see that the results of the posterior distributions of 
$\pr(\Delta^{(1)} | \ D_2^0,...,D_4^3)$ depend on the prior choice A and B, but only marginally on $\alpha_\eta$ and  $\beta_\eta$.
 As for the posterior distributions of the truncation uncertainty, we find that the truncation errors in $\mu^2$H do not change by much while in $\mu^4$He$^+$ they become larger and for $\mu^3$He$^+$ and $\mu^3$H they are reduced upon {\it combining} the data sets. Again, the calculated posterior distributions have much larger tails that the Gaussian distribution for all prior choices.
 
The $\eta$-prior $\alpha_\eta$ efficiently summarizes, through the maximum-entropy principle, the information available on the parameter $\eta$ before the calculation of the data sets. The analysis with prior $\beta_\eta$, on the other hand, makes additional assumptions about the value of $\eta$, \emph{e.g.}, it constrains $\eta$ to a value smaller than 1. While this assumption is well supported by both prior and posterior information available to us, it does not allow us to diagnose divergences of the expansion in Eq.~\eqref{eq: delta_pol_model}. Furthermore, the analysis with prior B for $\pr(c_i|\bar{c})$ is more sensitive to details of the distributions of the coefficients $c_i$ compared with the analysis with prior $A$. We therefore consider the uncertainty estimates obtained with the combinations of priors $\alpha_\eta$ and $B$ to be better calibrated than the other choices explored in this work.

\begin{table}[h!]
\begin{center}
\begin{tabular}{ c | c | c | c | c | c | c}
 \multirow{2}{*}{Atom} & \multirow{2}{*}{Prior} & $68 \%$ CI  & $68 \%$ CI  & $95 \%$ CI  & $95 \%$ CI  &  Ref.~\cite{Ji_2018} \\ 
~ & ~& $(\alpha_\eta)$ & $(\beta_\eta)$ & $(\alpha_\eta)$ & $(\beta_\eta)$&   \\
 \hline
 \multirow{2}{*}{$\mu^2$H} & A & 0.60 & 1.03 & 2.81 & 3.72 & \multirow{2}{*}{0.4}\\ 
 & B & 0.79 & 1.38 & 3.46 & 4.07 &\\
 & & & & & &\\
 \multirow{2}{*}{$\mu^3$H} & A & 1.18 & 1.73 & 5.44 & 6.67 & \multirow{2}{*}{1.3}\\ 
 & B & 1.52 & 2.29 & 6.88 & 7.73 &\\
 & & & & & &\\
 \multirow{2}{*}{$\mu^3$He$^+$} & A & 4.95 & 4.90 & 14.47 & 13.05 & \multirow{2}{*}{1.1}\\ 
 & B & 4.81 & 4.74 & 14.99 & 13.87 &\\
 & & & & & &\\
 \multirow{2}{*}{$\mu^4$He$^+$} & A & 0.63 & 1.02 & 3.32 & 4.86 & \multirow{2}{*}{0.8}\\ 
 & B & 0.89 & 1.55 & 4.70 & 5.80 &\\
\end{tabular}
\caption{ Bayesian analysis of the {\it separated} muonic systems: truncation uncertainties in the $\eta$-expansion of $\delta_{\text{pol}}^{A}$ expressed as  confidence interval $\%$ for various prior choices. }\label{tab: 68_95}
\end{center}
\end{table}

\begin{table}[h!]
\begin{center}
\begin{tabular}{ c | c | c | c | c | c | c}
 \multirow{2}{*}{Atom} & \multirow{2}{*}{Prior} & $68 \%$ CI  & $68 \%$ CI  & $95 \%$ CI  & $95 \%$ CI  & Ref.~\cite{Ji_2018} \\ 
~ & ~& $(\alpha_\eta)$ & $(\beta_\eta)$ & $(\alpha_\eta)$ & $(\beta_\eta)$&  \\
 \hline
 \multirow{2}{*}{$\mu^2$H} & A & 1.06 & 1.09 & 1.81 & 1.91 & \multirow{2}{*}{0.4}\\ 
 & B & 0.87 & 1.01 & 2.43 & 2.77 &\\
 & & & & & &\\
 \multirow{2}{*}{$\mu^3$H} & A & 1.12 & 1.17 & 1.94 & 2.05 & \multirow{2}{*}{1.3}\\ 
 & B & 0.93 & 1.09 & 2.69 & 3.14 &\\
 & & & & & &\\
 \multirow{2}{*}{$\mu^3$He$^+$} & A & 2.37 & 2.44 & 4.07 & 4.31 & \multirow{2}{*}{1.1}\\ 
 & B & 1.96 & 2.30 & 5.65 & 6.58 &\\
 & & & & & &\\
 \multirow{2}{*}{$\mu^4$He$^+$} & A & 2.18 & 2.25 & 3.74 & 3.95 & \multirow{2}{*}{0.8}\\ 
 & B & 1.79 & 2.07 & 4.84 & 5.42 &\\
\end{tabular}
\caption{  Bayesian analysis of the {\it combined} muonic systems: truncation uncertainties in the $\eta$-expansion of $\delta_{\text{pol}}^{A}$ expressed as  confidence interval $\%$ for various prior choices.
}
\label{tab: 68_95_comb}
\end{center}
\end{table}

We would like to comment on the fact that we get an $\eta$-expansion uncertainty  larger than previously estimated for $\mu ^3$He$^+$. As already pointed out  in Ref.~\cite{Nevo_Dinur_2016}, this muonic system does not display the expected scaling in $\eta$,  where, as one can see from  Table~\ref{tab: data}, $D_3$ is unusually small and $D_4$ unusually large.
In Ref.~\cite{Ji_2018}\footnote{Note that here relativistic terms were included.}, the $\eta$ parameter has been estimated by taking an average of the ratio between the first terms appearing in the Taylor expansion of Eq.~(\ref{eq: delta_pol_expansion}), namely  $\eta \sim 4|D_3/D_2|$ and $\eta \sim \sqrt{20}|D_4/D_2|^{1/2}$, and the final uncertainty has been calculated as $\frac{1}{120}\eta^3$, which is the first omitted term in Eq.~(\ref{eq: delta_pol_expansion}). 
Here, instead, we do not take any coefficients from the Taylor expansion, but  leave the 
 constants $c_i$  of Eq.~(\ref{eq: delta_pol_model}) free to float, requiring that they are  natural in our analysis. Hence,  the Bayesian analysis is on the one hand sensitive to the unusually large ratio $D_4/D_3$ and on the other hand insensitive to the suppressing factor $1/120$, which explains why we get a larger uncertainty.
 In general, the Bayesian analysis gives more conservative uncertainty estimates {\it for a given value of $\eta$}. However, since the posterior probability distributions of  $\eta$ for  $\mu^2$H, $\mu^3$H, and $\mu^4$He$^+$ peak below $\eta=0.33$, our 68$\%$ CI are in good agreement with the estimates from Ref.~\cite{Ji_2018}.
 For $\mu^3$He$^+$ the posterior probability distribution peaks slightly above 0.33  leading to larger uncertainties compared to  Ref.~\cite{Ji_2018}.

\section{Conclusion}
In this work, we have performed a Bayesian analysis of the polarizability data sets of the nuclear structure corrections to the Lamb shift in $\mu^2$H, $\mu^3$H, $\mu^3$He$^+$ and $\mu^4$He$^+$. For the $\mu^2$H, $\mu^3$H and $\mu^4$He$^+$ systems we find that the  maximum-likelihood value for $\eta$ is about half the value of $0.33$ estimated using the uncertainty principle. Despite the smaller value of the expansion parameter, the Bayesian analysis gives $68 \%$ CIs for the truncation uncertainty that are in good agreement with the estimates in Ref.~\cite{Ji_2018}. 
When compared to the other muonic systems, both the value of $\eta$ and the value of the truncation uncertainty in $\mu^3$He$^+$ are anomalously large. Most likely this is the consequence of the very large sub-sub-leading correction $(D_4)$ to $\delta^{\text{A}}_{\text{pol}}$ in this system. From our analysis, we find that the maximum-likelihood value of $\eta$ is 0.35, whereas the $68 \%$ CI of the truncation uncertainty is $\sim 5 \%$ of the total non-relativistic polarizability contribution. When {\it combining} the data sets we obtain an ``average" of the results coming from the individual analysis, with the uncertainties in $\mu^2$H being overall unchanged, the uncertainties in $\mu^4$He$^+$ becoming slightly larger, the uncertainties in $\mu^3$H slightly smaller and the uncertainties in $\mu^3$He$^+$ becoming considerably smaller. 

The analysis indicates that the $\eta$-expansion in $\mu^3$He$^+$ might converge slower than previously expected. Our updated value for the truncation uncertainty of the $\eta$-expansion in $\mu^3$He$^+$ is as large as other  contributions, such as the Coulomb term (see Table 8 in Ref.~\cite{Ji_2018}). A possible solution  could be found by using  the $\eta$-less method \cite{Hernandez19,Acharya:2020bxf} for the evaluation of the nuclear polarizability effects in $\mu^3$He$^+$, that completely avoids the expansion in $\eta$, but requires the more cumbersome calculation of the longitudinal and transverse response functions. This method has  been so far successfully implemented and used only for the $\mu^2$H system. Another solution may be the inclusion of the next term in the $\eta$-expansion of Eq.~(\ref{eq: delta_pol_expansion}). We can speculate that this approach could potentially  make the $\eta$-expansion uncertainties in $\mu^3$H and $\mu^4$He$^+$ negligible compared to the other uncertainty sources, while  in the case of $\mu^3$He$^+$ the $\eta$-expansion uncertainties may just become comparable to the previous estimates of Refs.~\cite{Ji_2018,Ji_2013}. 

Quantifying and reducing theory uncertainties in its various sources is important because improving the precision of the polarizability calculation will enable us to extract more precise values of charge radii than the current state of the art. In recent years a new set of nuclear interactions constructed at different orders in chiral-EFT has become available \cite{Gezerlis14,Tews16,Lynn16,Lynn17,LiMuli21}. This work, in conjunction with the new set of interactions, sets the stage for performing a study of the chiral-EFT truncation uncertainties in muonic atoms heavier than $\mu^2$H with Bayesian techniques. This activity in muonic atoms, combined with the ongoing experiments in Garching and Amsterdam in ordinary Helium ions \cite{Krauth19,Herrmann09} will in turn provide further tests of bound-state QED.

\noindent{\bf Acknowledgments} \\
We would like to thank Daniel Phillips, Nir Barnea and Chen Ji for useful discussions.  
This work was supported by the Deutsche Forschungsgemeinschaft (DFG) with the Collaborative Research Center 1044 and
through the Cluster of Excellence ``Precision Physics, Fundamental
Interactions, and Structure of Matter" (PRISMA$^+$ EXC 2118/1) funded by the
DFG within the German Excellence Strategy (Project ID 39083149). BA also acknowledges support by the Neutrino Theory Network Program (Grant No. DE-AC02-07CH11359).

\section*{References}
\bibliography{mybibfile.bib}

\end{document}